# Mixed cationic liposomes for brain delivery of drugs by the intranasal route: the acetylcholinesterase reactivator 2-PAM as encapsulated drug model

Tatiana N. Pashirova[a*], Irina V. Zueva[a], Konstantin A. Petrov[a,b], Svetlana S. Lukashenko[a], Irek R. Nizameev[a,c], Natalya V. Kulik[a], Aleksandra D. Voloshina[a], Laszlo Almasy[d], Marsil K. Kadirov[a,c], Patrick Masson[b], Eliana B. Souto[e,f], Lucia Ya. Zakharova[a], Oleg G. Sinyashin[a]

[a]*Arbuzov Institute of Organic and Physical Chemistry, FRC Kazan Scientific Center of RAS, Arbuzov St., 8, Kazan, 420088, Russia*

[b]*Kazan Federal University, Kremlyovskaya St., 18, Kazan, 420008, Russia*

[c]*Kazan National Research Technological University, Karl Marx St., 68, 420015, Kazan, Russia*

[d]*Institute for Solid State Physics and Optics, Wigner Research Centre for Physics, Hungarian Academy of Sciences, Budapest, Hungary*

[e]*Department of Pharmaceutical Technology, Faculty of Pharmacy, University of Coimbra, Coimbra, Portugal*

[f]*REQUIMTE/LAQV, Group of Pharmaceutical Technology, Faculty of Pharmacy, University of Coimbra, Coimbra, Portugal*

Keywords: Cationic Liposomes, Surfactant, Self-assembly, Pralidoxime chloride, Paraoxon, Drug Delivery Systems, Acetylcholinesterase, Organophosphorus agent

**Corresponding Author**

Tatiana N. Pashirova (Corresponding author)
A.E. Arbuzov Institute of Organic and Physical Chemistry, FRC Kazan Scientific Center of RAS, Arbuzov str. 8, 420088, Kazan, Russia
E-mail: tatyana_pashirova@mail.ru, tel.: +79871722060; fax: +78432732253

ABSTRACT: New mixed cationic liposomes based on L-α-phosphatidylcholine and dihexadecylmethylhydroxyethylammonium bromide (DHDHAB) were designed to overcome the BBB crossing by using the intranasal route. Synthesis and self-assembly of DHDHAB were performed. A low critical association concentration (0.01 mM), good solubilization properties toward hydrophobic dye Orange OT and antimicrobial activity against gram-positive bacteria *Staphylococcus aureus* (MIC=7.8 µg·mL$^{-1}$) and *Bacillus cereus* (MIC=7.8 µg·mL$^{-1}$), low hemolytic activities against human red blood cells (less than 10%) were achieved. Conditions for preparation of cationic vesicle and mixed liposomes with excellent colloidal stability at room temperature were determined. The intranasal administration of rhodamine B-loaded cationic liposomes was shown to increase bioavailability into the brain in comparison to the intravenous injection. The cholinesterase reactivator, 2-PAM, was used as model drug for the loading in cationic liposomes. 2-PAM-loaded cationic liposomes displayed high encapsulation efficiency (~ 90 %) and hydrodynamic diameter close to 100 nm. Intranasally administered 2-PAM-loaded cationic liposomes were effective against paraoxon-induced acetylcholinesterase inhibition in the brain. 2-PAM−loaded liposomes reactivated 12 ± 1% of brain acetylcholinesterase. This promising result opens the possibility to use marketed positively charged oximes in medical countermeasures against organophosphorus poisoning for reactivation of central acetylcholinesterase by implementing a non-invasive approach, via the "nose-brain" pathway.

# 1. Introduction

Intranasal route remains one of the elective non-invasiveness routes for the treatment of brain diseases [1]. It is highly appealing in clinical settings due to rapid central action of drug. This expands drug efficiency in emergency treatment [2]. The nasal mucosa allows direct delivery of drugs to the brain bypassing the blood brain barrier (BBB). It avoids hepatic clearance and drug degradation by liver enzymes, and therefore, increases the effect of neurotherapy [3]. Applicability of intranasal route is limited by lipophilicity and molecular weight of drugs [4]. New formulations to maximize the concentration of drugs and new devices for intranasal administration have been developed [5]. To improve drug transfer to the brain and to prolong the residence time of drug in the nasal cavity, several strategies may be employed, namely the use of (i) permeation enhancers, muco-adhesive and heat sensitive gels, or (ii) nano-sized drug carriers. Colloidal nanosized carriers are especially important for brain delivery and targeting of drugs for treatment of neurodegenerative diseases [6]. Intranasal delivery to the brain by using nanosized particles can be carried out for small molecules, as well as for proteins, hormones, vaccines, DNA, etc [7-10]. The main strategy to improve drug delivery to the brain by intranasal administration is the modification of nanoparticles by absorption and permeation enhancers. For that purpose, surfactants, bile salts, cyclodextrins, derivatives of chitosan, arginine, lactoferrin, have already been proposed [11-16]. Liposomes are well-known universal nano-carriers. Their surface can be easily modified to provide them with multiple functions [17]. This has already been demonstrated for improvement of intranasal delivery [18] of rivastigmine [19,20], donepezil [21], zolmitriptan [22].

The present work was aimed at using intranasal drug delivery as an alternative route to crossing of the BBB by using cationic liposomes as drug carriers. Cationic liposomes usually consist of natural neutral phospholipids and positively charged lipids [23]. Cationic liposomes deserve attention and further development since the ability of nanoparticles to penetrate the nasal mucus is obviously dependent not only on the size, but also, primarily, on the surface charge of particles [24]. In addition, the positive charge of liposomes plays an important role in the *in vitro* processes of interaction with cells epithelium [25] and as a mucosal adjuvant [8,26]. We applied the cationic liposome intranasal drug delivery approach in the treatment of organophosphorus poisoning. Organophosphorus agents (OPs) are highly toxic chemicals used as pesticides; some of them are banned chemical warfare agents. OPs represent a global chemical threat for both armed forces and civilian populations [27]. The acute toxicity of OPs results from inhibition of acetylcholinesterase (AChE). Acute poisoning causes death due to respiratory failure, seizures and irreversible brain damage. The current post-exposure treatment is to administer a combination of an AChE reactivator, atropine as an anticholinergic drug, and an anticonvulsant to prevent irreversible brain damage [27,28]. However, current AChE reactivators (quaternary oximes) are unable to reactivate phosphylated AChE in the central nervous system because they do not cross significantly the BBB [29]. Different strategies have been implemented for allowing these reactivators to cross the BBB [30]. In particular, non-quaternary oximes capable of crossing the BBB were synthesized. However, none of newly synthesized molecules has been developed so far [31,32]. New trends have emerged with the development of nanotechnology and new delivery systems, making quaternary oximes capable of crossing the BBB [33-35]. The intranasal route has also been successfully tested for administration of drugs against OP poisoning. For example, intranasal midazolam led to cessation of soman-induced seizures [36]. Intranasal administration of oximes has been considered as an alternative way for bypassing the BBB. Indeed, intranasal administration of obidoxime contributed to a reduction of brain damage and death of rats after poisoning by paraoxon [37]. Also, aerosolized butyrylcholinesterase, as a bioscavenger, was successful for protection of macaque lungs against volatile OPs [38]; rapid intranasal delivery of enzymes to the brain indicates that this route could be used for protection of CNS against toxicants [39], including OPs. Our present results validate the intranasal brain delivery approach using cationic liposome as drug carriers.

**2. Materials and methods**

**2.1. Chemicals.** Dihexadecyldimethylammonium bromide (DHDAB, 97%, Sigma-Aldrich), Tween® 80 (Polysorbate 80) (BioXtra, Sigma-Aldrich, product of France), L-α-phosphatidylcholine (Soy, 95%, Avanti polar lipids), Phospholipon® 80 and Lipoid® S75, 75% soybean phosphatidylcholine were gifts from Lipoid GmbH (Ludwigshafen, Germany), 1-(o-Tolylazo)-2-naphthol (75%, Orange OT, Aldrich, USA), pyrene (99%, Sigma, Switzerland), Pyridine-2-aldoxime methochloride (2-PAM chloride, Pralidoxime chloride) (≥ 97 %, Sigma-Aldrich, product of China). Dihexadecylmethylhydroxyethylammonium bromide (DHDHAB) was synthesized as described in Supplementary material. Ultra-purified water (18.2 MΩcm resistivity at 25°C) was produced from Direct-Q 5 UV equipment (Millipore S.A.S. 67120 Molsheim-France). All reagents were used without further treatment. DHDAB and DHDHAB are insoluble in water at room temperature. Their solutions were prepared by heating (up to 60 ºC) in water without sonication or other energy except a manual mixing. Stable dispersions of surfactants were observed during cooling to room temperature (25 ºC). Stock solutions prepared at the concentration 10 mM were subsequently diluted with pure water to final working solutions.

**2.2. Fluorescence.** Fluorescence spectra of pyrene ($1\times10^{-6}$ M) in water solutions of cationic surfactants were recorded at 35°C on a Varian Cary Eclipse spectrofluorimeter (Varian, Inc., California, USA) with excited wavelengths for pyrene at 335 nm using 1 cm path length quartz cuvettes. Emission spectra were recorded in the 350-500 nm range. The fluorescence anisotropy of 1,6-diphenyl-1,3,5-hexatriene (DPH) was measured on a Cary Eclipse (spectrometer (Varian, Inc., California, USA)) equipped with filter polarizers. DPH were excited at 361 nm, and the fluorescence intensity was measured at 450 nm. The embedded software automatically determined the correction factor and anisotropy value. A quartz cell of 1 cm path length was used for all fluorescence measurements. Measurements were performed in thermostated cell at 35°C. A fixed concentration of 0.175 mM of fluorescence probe DPH was used.

**2.3. Surface Tension.** Surface tension measurements were carried out using several methods: i) the du Nouy ring detachment method (Kruss K6 Tensiometer, Hamburg, Germany). Briefly, the planar and spherical platinum ring was placed parallel to the air/solvent interface ii) the Wilhelmy plate method (Kruss K9 Tensiometer, Hamburg, Germany). The platinum plate was wetted to a set depth. A contact angle θ of 0° (cos θ = 1) with liquids was formed. The surface tension was calculated from the measured force. Between two surface tension measurements, the platinum ring and the platinum plate were cleaned with Ultra-purified water, then soaked in ethanol for 5–7 min, rinsed again with Ultra-purified water, and finally flame-dried; iii) the bubble pressure method. In this method air gas bubble was blown at constant rate through a capillary tube submerged into the surfactant solutions. The surface tension was calculated by the Laplace equation *($\Delta P_{max} = 2\gamma/r$*, where *γ* is the surface tension, and *r* is the capillary tip radius, using the measured maximum pressure difference $\Delta P_{max}$; iv) the drop weight (stalagmometric) method [40]. For this purpose, several drops of liquid leaked out of the glass capillary of the stalagmometer were weighed. The surface tension was determined from the number of drops leaking out; v) The pendant drop shape analysis technique [41, 42] is one of the most accurate methods to measure surface tension; it has many advantages over other methods [43, 44]. The surface tension was measured with a DSA 30 (Kruss tensiometer) using a 2.5 mL syringe with a fitted steel capillary of 1.84 mm outer radius. The DSA 30 tensiometer allows recording of the drop shape with a built-in digital video camera. The theoretical drop shape determined from the Young–Laplace equation was fitted to experimental drop profiles. The measured surface tension corresponded to the best fit. Temperature was kept at 35 ± 0.2°C during all experiments.

**2.4. Atomic force microscopy.** An atomic force microscope (MultiMode V, USA) was used to reveal the morphology of the samples. RTESP cantilevers (Veeco, USA) with silicone tips were used in all measurements. The scanning rate was 1 Hz used to eliminate external distortions. Droplets of aqueous dispersions of the samples were carefully placed on mica surface. The AFM imaging was performed after water evaporation.

**2.5. Small-angle neutron scattering**. SANS measurements on DHDHAB solutions were performed on the Yellow Submarine instrument operating at the Budapest Research Reactor [45]. The surfactant was dissolved in $D_2O$ at concentrations 0.6 and 1.21 mg/ml, filled in planar quartz cells of 2 mm path length, and measured at 37± 0.5°C, over a $q$-range of 0.08 – 3 $nm^{-1}$. The scattering curves were modeled by form factors of unilamellar vesicles, using the software SasView.

**2.6. Solubilization study.** Solubilization study of the dye (Orange OT) was performed by adding an excess of crystalline Orange OT to the cationic surfactant solutions. These solutions were allowed to equilibrate for about 48 h at 35°C, followed by filtration, and the absorbance was measured at 495 nm using a spectrophotometer Specord 250 Plus (Analytik Jena AG, Germany). Quartz cuvettes (1 cm cell path) containing sample were used. Solubilization capacity of associates (*S)*, which corresponds to the number of moles of dye solubilized per mole of surfactant was determined according to equation 1 [46,47]

$$S = B/(\varepsilon_{ext} \times l) \quad (1)$$

where *B* is the slope of dye absorbance as a function of surfactant concentration above CAC and $\varepsilon_{ext}$ the extinction coefficient of Orange OT ($\varepsilon_{ext}$ = 18720 $M^{-1}$ $cm^{-1}$).

**2.7. Antibacterial and antifungal activity.** *In vitro* antibacterial and antifungal activities of the cationic surfactant were evaluated against pathogenic representatives of gram-positive and gram-negative bacteria and fungi. Minimal inhibitory concentrations (MICs) were estimated by conventional dilution methods for bacteria and fungi [48]. The antibacterial and antifungal assays were performed in Hottinger broth (HiMe-dia Laboratories Pvt. Ltd Mumbai, India) and Sabouraud dextrose broth (HiMedia Laboratories Pvt. Ltd Mumbai, India) (bacteria 3 × $10^5$ cfu/ml and yeast 2 × $10^4$ cfu/ml). The components of Hottinger broth (Lactalbumin hydrolysate (10 g), Peptone (10 g), NaCl (5 g)) were dissolved in one liter of distilled water, autoclaved for 20 min at 121°C. The pH was adjusted to 7.2 prior to autoclaving. The components of Sabouraud dextrose (Peptone (10 g), glucose (40 g)) were dissolved in one liter of distilled water, autoclaved for15 min at 121°C. The pH was adjusted to 5.6 prior to autoclaving.

**2.8. Hemolytic activity assay.** The toxicity of cationic surfactant was tested for their hemolytic activities against human red blood cells (hRBC). Fresh hRBC collected from heparinized blood were rinsed 3 times with 35 mM phosphate buffer/0.15 M NaCl, pH 7.3 by centrifugation for 10 min at 800 g and re-suspended in PBS. Test compound dissolved in PBS (concentrations 0.98–500 μg/ml) were then added to 0.5 mL of a stock solution of hRBC in PBS to reach a final volume of 5 mL (final erythrocyte concentration was 10% v/v). The resulting suspension was incubated under stirring for 1 h at 37°C. The samples were then centrifuged at 2000 rpm for 10 min. Release of hemoglobin from hRBC was monitored by measuring the supernatant absorbance at 540 nm. Controls for zero hemolysis (blank) and 100% hemolysis consisted of hRBC suspended in PBS and distilled water, respectively.

**2.9. Preparation of cationic DHDHAB vesicles and mixed cationic liposomes.** L-α-phosphatidylcholine and cationic surfactants (7% w/w) were dissolved in 2 ml of ethanol. The homogeneous solution was kept in a water bath at 60°C until alcohol evaporation to obtain a thin lipid film. Ultra-purified water (Milli-Q, Direct-Q5 UV) or saline (0.9%) was pre-heated to 60°C and added to rehydrate the lipids at 60°C in the absence or presence of 2-PAM (0.35% w/w). The solution was stirred under magnetic stirring (750 rpm) (Heidolph, Germany) for 30 min at the same temperature. Then the solution was kept for 1.5 hours in a water bath at 37°C. The multilamellar liposomes were extruded 20 times by passage through a polycarbonate membrane of 100 nm pore size (Mini-Extruder Extrusion Technique, Avanti Polar Lipids, Inc.). For preparation of cationic vesicles of DHDHAB at concentration up to ca. 1.0 mM the same process was performed. For the loading of liposomes with rhodamine B, the same method was used by replacing 2-PAM with rhodamine B (0.01% w/w).

**2.10. Morphology and particle size analysis.** The mean particle size, zeta potential and polydispersity index of nanoparticles were determined by dynamic light scattering (DLS) measurements, using the Malvern Instrument Zetasizer Nano (Worcestershire, UK) at neutral pH (close to 7). Measured autocorrelation functions were analyzed by Malvern DTS software, applying the second-order cumulant expansion methods. The effective hydrodynamic radius ($R_H$) was calculated according to the Einstein-Stokes equation $D=k_BT/6\pi\eta R_H$, where $D$ is the diffusion coefficient, $k_B$ the Boltzmann constant, $T$ the absolute temperature, and $\eta$ the viscosity. The diffusion coefficient was measured at least three times for each sample. Five measurements of electrophoretic mobilities on each sample were converted into zeta potential by using the Smoluchowski relationship [49]: $\zeta=\mu\eta/\varepsilon$, where $\zeta$ is the zeta potential, $\eta$ the dynamic viscosity of the fluid, $\mu$ the particle mobility and $\varepsilon$ the dielectric constant. All samples were diluted with ultra-purified water to suitable concentration and analyzed in triplicate. Analysis of the optimal formulation was performed by fluorescence microscopy (Leica DM IRBE/Leica TCS SP2) to confirm the presence of liposomes, to verify the shape of the particles, and the absence of aggregation.

**2.11. Encapsulation efficiency and loading capacity.** Encapsulation efficiency (EE, %) and loading capacity (LC, %) were assessed for samples containing 2-PAM and Rhodamine B. These parameters were determined indirectly by filtration/centrifugation technique, measuring free 2-PAM (non-encapsulated) by spectrophotometry. A volume of 0.5 mL of each 2-PAM-loaded cationic liposomes was placed in centrifugal filter devices Ultracel 100K (100.000 MWCO, Amicon Millipore Corporation, Bedford, Massachusetts) to separate lipid and aqueous phases and centrifuged at 10000 rpm, for 15 minutes (Eppendorf AG, Hamburg, Germany). For liposomes containing rhodamine B, the same method was used, 2-PAM was replaced by Rhodamine B. Free 2-PAM and Rhodamine B were quantified by UV absorbance using a Specord 250 Plus (Analytik Jena AG, Germany) at 294 nm for 2-PAM (the extinction coefficient of 2-PAM at 294 nm is 11962 $M^{-1}\cdot cm^{-1}$ in 0.025 M phosphate buffer at pH = 7.4) and at 555 nm for Rhodamine B (the extinction coefficient of Rhodamine B at 555 nm is 92330 $M^{-1} \cdot cm^{-1}$ in 0.0025 M phosphate buffer at pH = 7.4) The parameters were calculated against appropriate calibration curve, using the following equation [34]:

$$EE(\%) = \frac{Total\ amount\ of\ 2PAM - Free\ 2PAM}{Total\ amount\ of\ 2PAM} \times 100\%  \quad (2)$$

$$LC(\%) = \frac{Total\ amount\ of\ 2PAM - Free\ 2PAM}{Total\ amount\ of\ lipid} \times 100\%  \quad (3)$$

*2.12. In vitro* **2-PAM release profile.** The monitoring of 2-PAM release from cationic liposomes was performed using the dialysis bag diffusion method. Dialysis bags retain liposomes and allow the released 2-PAM to diffuse into the dissolution medium. The bags were soaked in Milli-Q water for 12 h before use. One milliliter of cationic liposomes was poured into the dialysis bag and the two ends were sealed with clamps. The bags were then placed in a vessel containing 200 mL of 0.025 M sodium phosphate buffer pH 7.4, the receiving phase. The vessel was placed in a thermostatic shaker (New Brunswick, USA), at 37ºC, under a stirring rate of 150 rpm. At predetermined time intervals, 1 mL samples were withdrawn, and and their absorbance at 294 nm was measured using a Specord 250 Plus (Analytik Jena AG, Germany). All samples were analyzed in triplicate.

**2.13. Histology Analysis of Brain.** All experiments involving animals were performed in accordance with the guidelines set forth by the European Communities Council Directive of November 24, 1986 (86/609/EEC) and the protocol of experiments approved by the Animal Care and Use Committee of Kazan Federal University. Wistar rats of both sexes were purchased from the Laboratory Animal Breeding Facility (Branch of Shemyakin-Ovchinnikov Institute of Bioorganic Chemistry, Puschino, Moscow Region, Russia) and were allowed to acclimate to their environment in the vivarium for at least 1 week before experiments. Animals were kept in sawdust-lined plastic cages in a well-ventilated room at 20–22 °C in a 12-h light/dark cycle, 60−70% relative humidity and given *ad libitum* access to food and water. Free rhodamine B or rhodamine B loaded cationic liposomes were intravenously injected or administrated intranasally to Wistar rats at the same dose, 0.5 mg/kg. Non-treated animals were used as controls with respect to background fluorescence. 45 minutes after injection, animals were anesthetized, their brain was isolated and stored at −80°C. Frozen tissue was moved to a −20°C freezer 24 h before slicing. Tissues were sliced into 10 μm sections using Sakura Tissuestek Cryo3 microtome (Sakura Finetek, Torrance, CA). The tissue sections were examined with a Leica TCS SP5 confocal microscope (Leica, Germany) using a cyanine 3 (CY3) filter (excitation wavelength 550 nm and emission at 570 nm). Microphotographs were taken using the Leica LAS AF program.

**2.14. Measurement of *in vivo* brain AChE inhibition and reactivation level.** For *in vivo* brain AChE inhibition assay rats were poisoned by $0.8 \times LD_{50}$ of POX (600 μg/kg, *i.p.*). Whole brains were collected 1 h after POX administration and frozen in liquid nitrogen. Whole brains of rat control group were collected 1h after intraperitoneal (*i.p.*) injection of physiological saline. Thereafter, AChE activity of brain homogenates was analyzed. The mean brain AChE activity of poisoned animals was compared with mean brain AChE activity of control group.

For *in vivo* brain AChE reactivation assay, 1 h after poisoning by $0.8 \times LD_{50}$ of POX, 2-PAM in cationic liposomes was administered intranasally at the dose 7 mg/kg. Whole brains were then collected 1h after administration of 2-PAM- cationic liposomes and frozen in liquid nitrogen. The mean activity of brain AChE after intranasal administration of 2-PAM-cationic liposomes was compared to the mean brain AChE activity of poisoned group.

Whole brain homogenates were prepared in a Potter homogenizer with 0.05 M Tris-HCl, 1 % Triton X-100, 1 M NaCl, 2 mM EDTA; pH 7.0 (1 volume of brain for 2 volumes of buffer) at 4°C. The homogenates were centrifuged (10,000 rev/min, at 4°C) for 10 minutes using Eppendorf 5430R centrifuge with FA-45-30-11 rotor (Eppendorf AG, Hamburg, Germany).

For AChE activity assay, 50μl supernatant was incubated with 5μl of tetra-isopropyl pyrophosphoramide (iso-OMPA) – as a butyrylcholinesterase specific inhibitor - in final concentration 0.1 mM, for 30 minutes. Then, AChE-catalyzed hydrolysis ~~reaction~~ of substrate was started by adding of 10 μl of acetylthiocholine iodide (final concentration 1 mM). After 10, 20 or 30 min samples reaction with substrate at 25°C, reactions were stopped by adding the carbamylating agent neostigmine (0.1 mM). Samples were diluted in 50 mM phosphate buffer (pH 8.0) and DTNB (0.1 mM) was added. AChE activity was measured according to the Ellman method[37] by determining the production of yellow 5-thio-2-nitro-benzoate anion, resulting from reduction of DTNB by thiocholine (the product of substrate hydrolysis) at 412 nm by spectrophotometry (PerkinElmer λ25). The amount of thiocholine produced during 20 min (10th–30th min) was calculated. Sample without substrate was used as a blank. All measurements for each sample were done in triplicate. AChE activity was expressed in relation to the amount of total protein, which was determined by the Bradford method. Data were analyzed using Origin 8 software and expressed as the mean ± SEM. Statistical analysis was performed using the Mann-Whitney test. P< 0.05 were considered statistically significant.

## 3. Results and discussion

In the present work, two cationic surfactants dihexadecyldimethylammonium bromide (DHDAB) and dihexadecylmethylhydroxyethylammonium bromide (DHDHAB) (Fig. 1) were used for modification of liposomes. Our interest focused on the properties of cationic surfactants with double, long-chains due to the fact that these surfactants are synthetic vesicle-forming amphiphiles capable of self-assembling spontaneously as giant uni-lamellar vesicles [50-52]. Properties and dissolution of surfactants strongly depend on temperature, the chain length of alkyl radicals and the type of counter ions [53]. Not enough attention has been paid to the influence of structure of polar head-group of such cationic surfactants on their properties. In this work main attention is paid to the effect of hydroxyl group in polar head-group of cationic surfactants with double, hexadecyl chains.

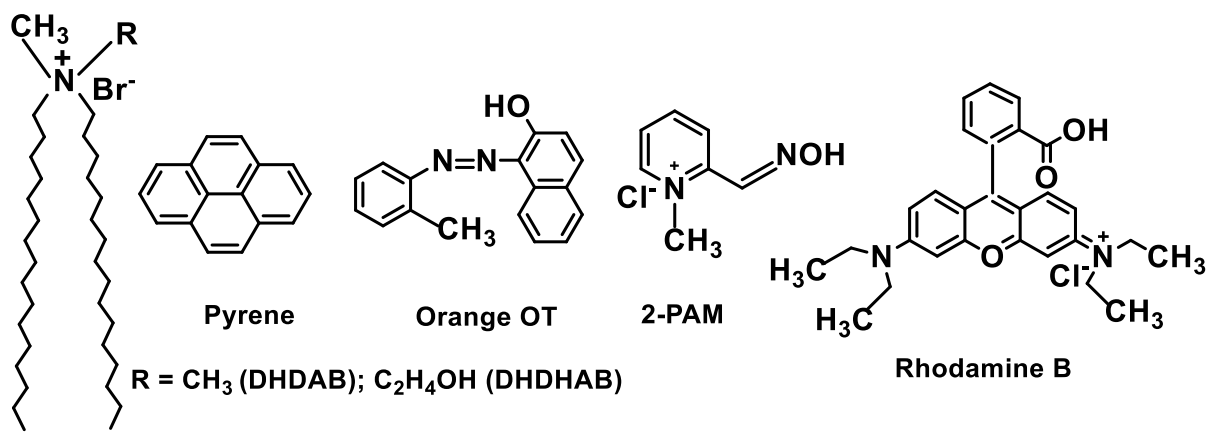

**Figure 1**. Structures of cationic surfactants dihexadecyldimethylammonium bromide (DHDAB) and dihexadecylmethylhydroxyethylammonium bromide (DHDHAB), hydrophobic dye (Orange OT) and fluorophore probes (Pyrene, Rhodamine B) and AChE reactivator (2-PAM).

*3.1. Self-assembly and solubilization properties of cationic surfactants*

Self-assembly of cationic surfactants was investigated by fluorescence, tensiometry, spectrophotometry and dynamic light scattering at 35°C.

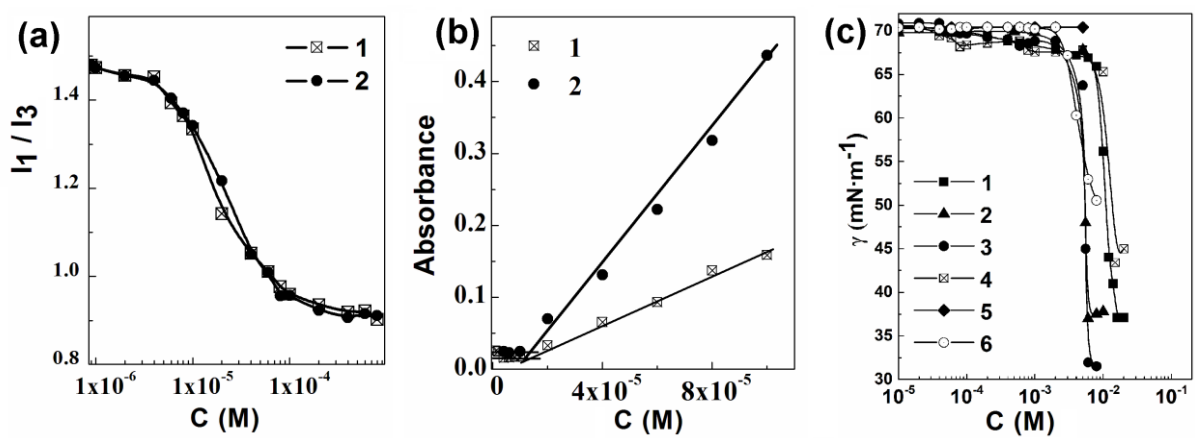

**Figure 2.** Dependence of the intensity ratio ($I_1/I_3$) of the first and third peaks of pyrene (a), optical density of Orange OT (b) on concentrations of DHDAB (1) and DHDHAB (2) and surface tension (c) on concentrations DHDHAB, using different techniques: Du Nouy Ring method with fresh solutions (1), the same method with solutions after 1 week storage (2); the Wilhelmy plate method (3), the bubble pressure method (4), the drop weight (stalagmometric) method (5), the pendant drop shape analysis (6); $C_{pyrene} = 1 \times 10^{-6}$ M, T = 35°C.

To determine the critical association concentration (CAC) of DHDAB and DHDHAB a technique with pyrene as fluorophore probe was used [54,55], which involved the analysis of the intensity of pyrene peaks at 373 nm ($I_1$) and 384 nm ($I_3$) (Fig 1S, Supplementary material). The ratio of the intensity of the first and third peaks ($I_1/I_3$) indicates the effect of microenvironmental polarity on pyrene fluorescence. The value ($I_1/I_3$) is decreasing with increasing surfactant concentration (Fig. 2a). This indicates the onset of surfactant aggregation and transfer of pyrene from the water phase to a nonpolar area of aggregates. After that value ($I_1/I_3$) remains unchanged. The obtained CACs for DHDAB and DHDHAB are similar, close to 0.025 mM.

The formation of associates at low concentrations was confirmed by solubilization of hydrophobic dye (Orange OT) (Fig 2b). The increase in optical density of Orange OT in cationic surfactant solutions was observed above the CAC1. This is due to solubilization of the dye in the hydrophobic area of surfactants associates. It should be noted that such kind of surfactants (dioctadecyldimethylammonium bromide) were also shown to be effective solubilizer /or colloidal stabilizer for the hydrophobic drug miconazole [56]. The CAC1s for both DHDAB and DHDHAB are 0.01 mM, which is close to the CAC1s determined by fluorescence. The solubilization capacity, determined by equation 1, is 0.06 and 0.22 for DHDAB and for DHDHAB, respectively. These values are 4 and 14 times, respectively, higher than the solubilization capacity of the classical surfactant CTAB.

Measurements of surface activity of DHDHAB at air-water interface using different techniques are presented in Fig. 2c. The decrease in surface tension ($\gamma$, mN·m$^{-1}$) is observed with increasing the surfactant concentration. It can be seen in Fig. 2c that the different techniques showed consistent results: the surface tension is reduced only at high concentration (close to 10 mM). A similar effect was observed for dioctadecyldimethylammonium bromide (DODAB) by Feitosa and Brown [57]. Critical concentration attributed to spontaneous formation of vesicles which was detected by optical spectroscopy and fluorescence at lower concentrations was not observed by tensiometry. It appears that physical and morphological properties, and CAC of double-tail cationic surfactant depend not only on concentration, solvent conditions, and the presence of additives, but also the sample preparation protocol and the history of temperature treatment [58,59]. For example, the CAC of DHDAB determined by Tucker was 0.01 mM [60] However, the CAC value found by Nagarajan is lower (0.0034 mM) [61]. Lower CAC (close to 0.0001 mM) were also determined using a linear relationship between log(CAC) and the number of carbon atoms in the dialkyl chain (n) [62] and CAC of DDAB [63-65]. These strongly different aggregation concentrations, observed by tensiometry and fluorescence probes can be explained by the difference of the utilized measurement techniques. The majority of techniques for measurement of adsorption micelle-forming surfactants at the water-air surface are not sensitive enough for adsorption of surfactants at the very low concentrations, when the surfactant molecules do not cover the entire surface. Indeed, unlike classical surfactants, the two alkyl-chain surfactants diluted from mother liquors to low concentrations likely form stable spontaneous vesicles rather than individual surfactant molecules capable to adsorb at the interface. Importantly, unlike fluorimetry and dye solubilization, the surface tension measurement method is aimed at the examination of the surface activity of amphiphilic molecules rather than bulk aggregation. It detects well the formation of a surfactant monolayer on the water-air interface, while the occurrence of the water-surrounded surfactant bilayers or vesicles in the bulk of the solution does not necessarily show up as surface effect. Therefore, surface tension remains insensitive to the presence of vesicles, while the light scattering and dye solubilization methods detect precisely aggregates in the bulk that are sufficiently large, or possess distinct hydrophobic compartment.

Size (hydrodynamic diameter) of DHDAB and DHDHAB associates was determined by dynamic light scattering at different surfactant concentrations. The size of nanoparticles about 500 nm was observed for DHDAB concentration $\geq$CAC1 (Fig 2S). The zeta potential of particles was + 60±2 mV, the polydispersity index was 0.5±0.02. All parameters did not change with increasing DHDAB concentration (Table 1S). In the case of DHDHAB, two different size distribution patterns were observed depending on the surfactant concentration. Population of 200 nm occurs at concentrations ≥ CAC1, while two populations, 30 nm and 200 nm were observed at DHDHAB concentrations ≥ 18 mM (Fig. 3S, Table 1S). The polydispersity index increased from 0.3 to 0.5 with increasing concentration. The fluorescence anisotropy of 1,6-diphenylhexatriene (DPH) was investigated for determination of the morphology of DHDHAB associates. In Fig 4S in Supplementary material, it can be seen that the fluorescence anisotropy of DPH is ≥0.1 in the concentration range of DHDHAB 0.01-0.03 mM. These values are in good agreement with spontaneous formation of vesicles. The anisotropy fluorescence of DPH is ≤0.1 at higher concentrations of DHDHAB (above 1mM). This value of anisotropy indicates the formation of micellar associates at high concentrations.

The structure of the DHDHAB associates was further analyzed by small-angle neutron scattering (SANS) [66]. The scattering curves show the characteristic shape of flat structures (within the resolution range of the observation window 0.1-10 nm), which can be identified as large unilamellar vesicles (Fig. 3).

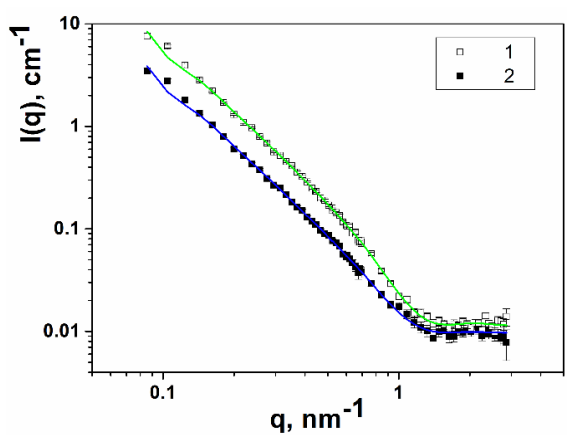

**Figure 3.** SANS scattering curves of dihexadecylmethylhydroxyethylammonium bromide at two different concentrations in D$_2$O at T = 37ºC: C$_{DHDHAB}$ = 2 mM (1), 1 mM (2).

Small micelles, even if present, could not be detected due to their much weaker scattering signal. In model fitting, the vesicle diameter and polydispersity was fixed at values 60 nm and 0.2, provided by light scattering, since the whole vesicle shape cannot be resolved at the given resolution limit. Model fitting gave membrane thickness of 4.14±0.08 nm and 4.13±0.5 nm for 1 and 2 mM DHDHAB solutions, respectively, which are in accord with the size of the surfactant molecule. The above presented results show that addition of hydroxyl group to the structure of the head group of surfactants with double hexyl chains can make a significant contribution both to self-assembly and solubilization properties of the investigated cationic surfactant dihexadecylmethylhydroxyethylammonium bromide.

*3.2. Antimicrobial activity and hemolysis of DHDHAB*

Determination of cytotoxicity or cell viability remains the most frequent test used as proof of biocompatibility of nanoparticles. It was reported that cytotoxicity of cationic liposomes composed of phosphatidylcholines and dialkyldimethylammonium bromide is increased with increasing surfactant concentration and their hydrophobic chain lengths [67]. Such kind of nanoformulations was recommended for topical therapy [68]. Cationic surfactants with double, long-chains exhibited antimicrobial properties [69]. It was confirmed that the determining factor of the antifungal action of cationic surfactants is not cell lysis, but a change in the cell surface from negative to positive charge [70]. Antimicrobial activity of DHDHAB surfactant is presented in Table 1. It is obvious that the DHDHAB surfactant is very effective against gram-positive bacteria *Staphylococcus aureus* (MIC=7.8 µg·mL$^{-1}$) and *Bacillus cereus* (MIC=7.8 µg·mL$^{-1}$). In order to provide evidence of biocompatibility of cationic surfactant formulations we investigated their action on hemolysis. The hemolytic action of DHDHAB surfactant on human erythrocytes is less than 10% in the concentration range used for liposomes surface modification.

**Table 1.**
Antimicrobial activity (mg/L) and haemolysis of human red blood cells (%)[a]

| Compound | Bacteriostatic and fungistatic activity (MIC), µg·mL$^{-1}$ | | | | | | | C / H[b] |
|---|---|---|---|---|---|---|---|---|
| | *Sa* | *Bc* | *E ci* | *Pa* | *Tm* | *An* | *Ca* | |
| DHDHAB | 7.8±0.6 | 7.8±0.6 | >500 | >500 | >500 | >500 | >500 | 7.8 / 5.5 |
| | Bactericidal and fungicidal activity, µg·mL$^{-1}$ | | | | | | | |
| | 31.3±2.8 | >500 | 31.3±2.7 | 250±22 | >500 | >500 | >500 | |

[a]The tests were performed in duplicate and repeated twice; [b]C is the concentration of DHDHAB (µg.mL$^{-1}$); H is haemolysis of human red blood cells (%).
*Sa, Staphylococcus aureus* ATCC 209P; *Bc, Bacillus cereus* ATCC 8035; *Pa, Pseudomonas aeruginosa* ATCC 9027; *Ec, Escherichia coli* CDC F-50; *An, Aspergillus niger* BKMF-1119; *Tm, Trichophyton mentagrophytes* var. *gypseum* 1773; *Ca, Candida albicans* ВКПГу-401/NCTC 885-653.

*3.3. Preparation and characterization cationic DHDHAB vesicles and cationic mixed Ph/DHDHAB liposomes*

Initially, the liposome preparation technique was developed using natural, commercially available and widely used phospholipids (Phospholipon 80, Lipoid S 75 and L-α-phoshatidylcholine (95%) and nonionic surfactant (Tween 80). The macroscopic appearance of each sample indicated that the formation of homogeneous multilamellar liposomes occurs only from L-α-phoshatidylcholine (Ph). We continued to improve the liposome solution by varying the concentration of the different components, their ratio, using several techniques (ultrasonic bath, ultrasonic probe and extrusion). The physicochemical characterizations of nano-formulations are presented in Table 2. The same liposome preparation technique was used for the preparation of DHDHAB cationic vesicles. The morphology and size of prepared DHDHAB cationic vesicles were investigated by atomic force microscopy. Small nanoparticles of spherical-ellipsoidal shape were observed. Nanoparticle diameter is 50 nm and their elevation is about 1-2 nm (Fig. 4) AFM results are comparable with properties measured by DLS (Table 2, Fig. 5S).

**Table 2.**
Hydrodynamic diameter (number, volume and intensity − averaged), polydispersity index (PdI) and Zeta potential (ZP) as a function of lipid structure, concentration and component ratio in different formulations, pH=7, t=25ºC.

| № | Formulation | Ratio Ph/surf., wt% | Loaded compound | hydrodynamic diameter, nm | | | PdI | ZP, mV |
|---|---|---|---|---|---|---|---|---|
| | | | | Number | Volume | Intensity | | |
| 1 | Ph/Tween | 99.5/0.5 | - | 68±2 | 91±2 | 122±1 | 0.25±0.01 | -14±2 |
| 2 | DHDHAB | 100 | - | 79±1 | 91±1 | 122±1 | 0.06±0.01 | - |
| 3 | Ph/DHDAB | 99.5/0.5 | - | 79±3 | 91±3 | 122±6 | 0.23±0.02 | +39±1 |
| 4 | Ph/DHDHAB | 99.9/0.1 | - | 91±3 | 122±2 | 164±3 | 0.143±0.01 | -1±0.3 |
| 5 | Ph/DHDHAB | 99.75/0.25 | - | 80±3 | 91±3 | 122±6 | 0.16±0.03 | +15±2 |
| 6 | Ph/DHDHAB | 99.5/0.5 | - | 91±3 | 106±2 | 122±6 | 0.1±0.03 | +25±2 |
| 7 | Ph/DHDHAB | 99/1 | - | 68±2 | 79±1 | 122±6 | 0.12±0.01 | +32±1 |
| 8 | Ph/DHDHAB | 95/5 | - | 79±1 | 106±2 | 164±3 | 0.16±0.02 | +64±2 |
| 9 | Ph/DHDHAB | 99.5/0.5 | Rhod B | 80±2 | 91±3 | 122±6 | 0.34±0.03 | +24±2 |
| 10 | Ph/DHDHAB | 99.5/0.5 | 2-PAM | 80±2 | 91±3 | 142±2 | 0.2±0.03 | +6±0.2 |

The conversion to intensity, number and volume-averaged distribution highlights different sizes of extruded vesicles with lower polydispersity index (0.06). DHDHAB extruded vesicles are not ideally spherical in accordance with AFM data. The size of DHDHAB cationic vesicles does not change during six months and more. This indicates the high stability of samples.

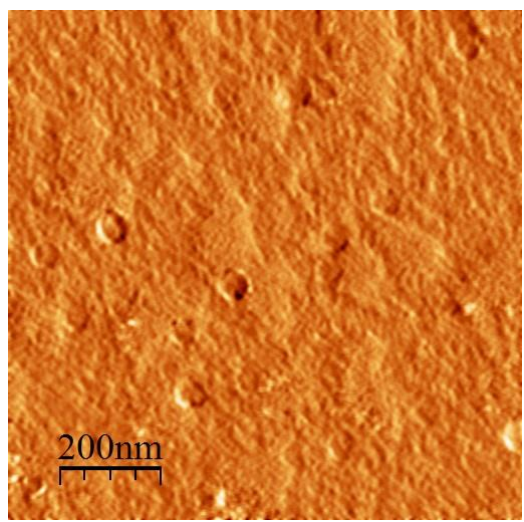

**Figure 4.** AFM image of DHDHAB cationic vesicles.

Double-tail surfactants were used for the preparation of mixed Ph/surfactant cationic liposomes. They differ considerably in lipid composition and surface charge. It is well known that the vesicle shape depends on membrane curvature and bending elasticity. Several experimental results indicate that various conditions (different physical stimuli and additives) can affect the vesicle shape [71-74]. Until now, a few examples have been established to evaluate the effect of a surfactant (varying in charge, length of carbon chain, structure of head group) on the properties of liposomes. There are a few examples showing that stability of cationic liposomes depends on the concentration and nature of cationic surfactants. It was shown that the adjuvantity of cationic liposomes also depends on the content of surfactant [75,76]. The addition of cationic surfactant (DHDAB and DHDHAB) did not provoke structural changes in liposomes morphology. The size of extruded cationic liposomes was close to extruded vesicles and their polydispersity was low (around 0.1). It was established that when cationic surfactants (DHDAB and DHDHAB) were used instead of a nonionic surfactant (Tween 80), the charge of mixed liposomes changed from negative (-14 mV) to positive (+36mV and +25 mV, respectively). Unlike Ph/Tween liposomes mixed cationic liposomes remain stable. An increase of the charge of cationic liposomes from -1 mV to +64 mV is possible by increasing the concentration of cationic surfactant from 0.1% to 5 % (w/w), respectively (form. № 4-8). The charge of mixed cationic Ph/DHDAB and Ph/DHDHAB liposomes is changed to -9mV and +6 mV respectively, in 6 months. The size of Ph/DHDAB liposomes decreased and polydispersity increased during the same time compared to Ph/DHDHAB liposomes. Likely, the addition of cationic surfactant (DHDAB) provoked structural changes in liposome morphology in course of time. Thus, Ph/DHDHAB liposomes were more stable during time.

*3.4. In vitro release studies and 2-PAM loading*

Cationic Ph/DHDHAB liposomes with composition 99.5/0.5 % (w/w) (form. №6) were chosen for encapsulation of pralidoxime chloride (2-PAM). The size of the 2-PAM-loaded liposomes was about 80 nm, the polydispersity index $0.2 \pm 0.03$, the zeta potential $+ 4 \pm 2$ mV. 2-PAM has a permanent positive charge, and in addition, at pH 7.4, about 50% of the oxime function is in oximate form. Thus, the decrease in zeta potential of cationic Ph/DHDHAB liposomes is likely due to positively charged and zwitterionic properties of 2-PAM. Encapsulation efficiency of 2-PAM, determined from equation 2, is rather high and equals almost 90%. Loading capacity of 2-PAM, determined from equation 3, is 22.5%. The dialysis method was performed to study the release of 2-PAM *in vitro* conditions. The release of 2-PAM from aqueous solutions and liposomes are presented in Fig. 5.

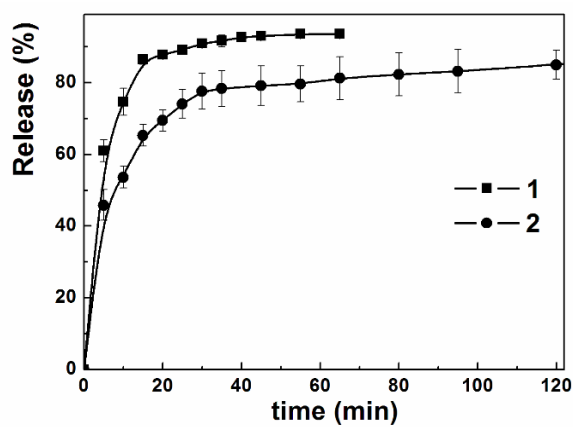

**Figure 5**. *In vitro* 2-PAM release from cationic Ph/DHDHAB liposomes using the dialysis bag method (n=3). 1- 2-PAM aqueous and 2- 2-PAM-liposomes solutions, phosphate-buffer (0.025 M), pH=7.4, 37°C.

The release of 2-PAM (90%) from the cationic Ph/DHDHAB liposomes takes three hours. It is three times slower than in the case of an aqueous solution.

*3.5. Intranasal administration of the fluorescent-labeled cationic liposomes*

To confirm the uptake of cationic Ph/DHDHAB liposomes into the brain tissue, cerebral cortex slices were analyzed by fluorescence microscopy (Fig. 6). Free rhodamine B was used as control. Physico-chemical characteristics of cationic mixed Ph/DHDHAB liposomes with included fluorescent probe (Rhodamine B) are presented in Table 2 (form. № 9). Encapsulation of Rhodamine B did not change both size and charge of cationic liposomes. EE (%) and LC (%) parameters determined from equations 2 and 3 are 98±1% and 1.4±0.02%, respectively. In Fig. 6 it is clearly seen that the greatest absorption of rhodamine B in brain occurs through the nasal route.

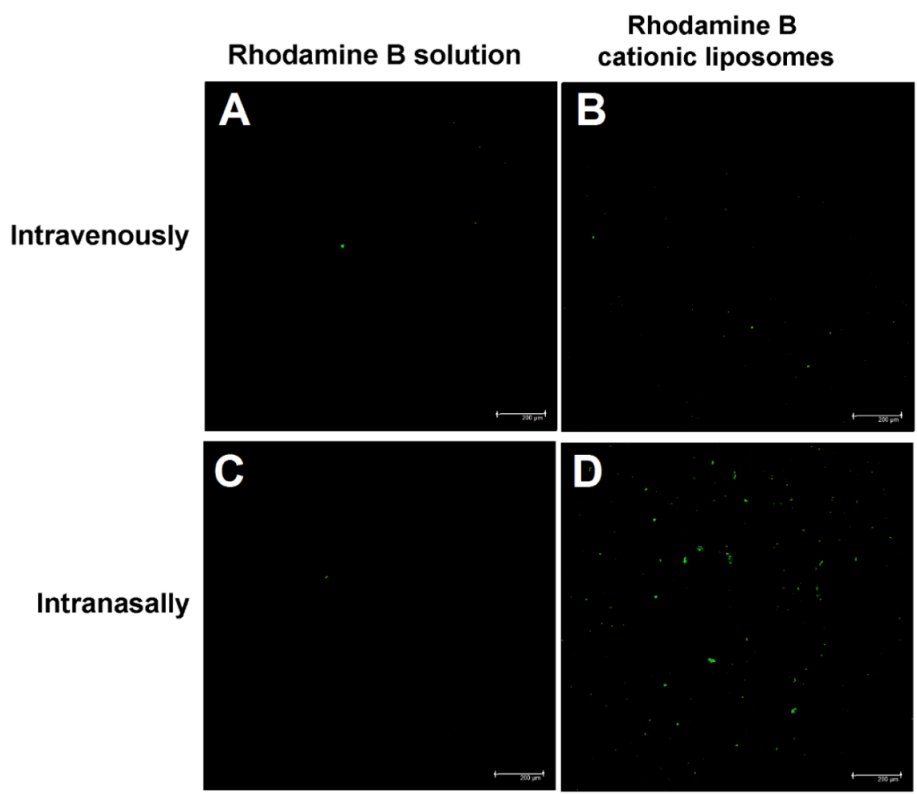

**Figure 6.** Section of cerebral cortex of rat treated with rhodamine B solution (A, C), treated with rhodamine B labeled cationic Ph/DHDHAB liposomes (B, D) by intravenous administration (A, B) and using intranasal route (C, D).

*3.6. Reactivation of Brain AChE in Vivo*

To test the *in vivo* efficacy of 2-PAM-loaded cationic Ph/DHDHAB liposomes against OPs, we used a rat model for OP toxicity using the organophosphate paraoxon (POX) as acetylcholinesterase inhibitor. After 1 h of poisoning by 0.8 × LD50 of POX (600 mg/kg, *ip*), brain samples were collected and AChE activity was analyzed. One hour after administration, this sub-lethal dose of POX (600 mg/kg, *ip*) resulted in 83.5 ± 4% inhibition of AChE in the brain (Figure 7). One hour after POX (600 mg/kg, *ip*), 2-PAM-loaded cationic liposomes were administered intranasally at the dose of 7 mg/kg. One hour after administration of 2-PAM-loaded cationic liposomes, activity of AChE was analyzed. It was previously found that intranasal administration of free 2-PAM (10 mg per animal) was ineffective in reducing paraoxon-mediated AChE inactivation in rat brain homogenates [37], whereas 2-PAM−liposomes reactivated 12 ± 1% of brain AChE (Figure 7). For substantial increase of AChE reactivation, more effective oximes could be used. However, gain in AChE reactivation level will certainly be obtained with 2-PAM by optimizing cationic surfactant formulations.

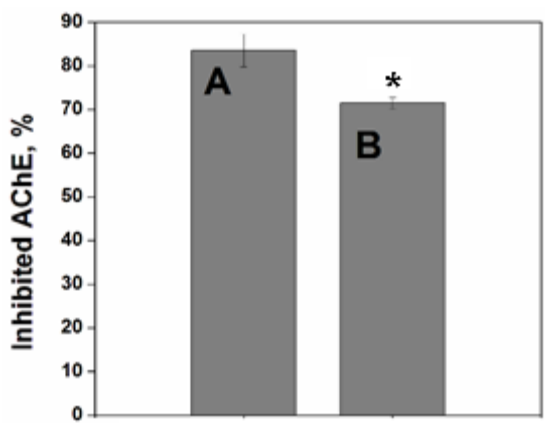

**Figure 7.** Determination of brain AChE reactivation level *in vivo*; mean AChE activity of brain homogenates was measured in a control group of rats (taken as 100%), after poisoning by $0.8 \times$ LD50 of POX (6 mg/kg, *ip*) (A) and intranasal administration of 2-PAM-loaded cationic liposomes (B), dose of 2-PAM 7 mg/kg. Data are presented as mean ± SE. (n = 5). * $p<0.05$ indicate significant difference by Mann–Whitney test.

**4. Conclusions**

The bioavailability of rhodamine B-loaded cationic liposomes to the brain when delivered by intranasal route was shown to be higher than when administered intravenously. 2-PAM-loaded cationic liposomes (the dose of 2-PAM was 7 mg/kg) intranasally administered to rat challenged with paraoxon ($0.8 \times$ LD50) were shown to reactivate central AChE (12%). This study provides evidence that reactivation of central AChE can be achieved by a non-invasive approach, i.e. to deliver oxime-loaded cationic liposomes to the brain via the "nose-brain" pathway.

**Acknowledgements**

**Author Contributions**

Tatiana N. Pashirova developed self-assembly and solubilization properties of cationic surfactants, preparation and characterization of cationic vesicles, mixed Ph/surfactant liposomes and performed *in vitro* drug release experiment; Irina V. Zueva performed biochemical experiments, histological analysis of brain; Konstantin A. Petrov performed analysis of brain sections by confocal microscopy; Svetlana S. Lukashenko synthesized and characterized cationic surfactant dihexadecylmethylhydroxyethylammonium bromide; Irek R. Nizameev performed the experimental work on atomic force microscope; Natalya V. Kulik tested the hemolytic activity against human red blood cells for cationic surfactant; Aleksandra D. Voloshina evaluated the *in vitro* antibacterial and antifungal activities of cationic surfactant; Laszlo Almasy performed the neutron scattering measurements and analysis; Marsil K. Kadirov analyzed data of atomic force microscopy; Patrick Masson proposed the idea of using of the intranasal way and wrote the manuscript; Eliana B. Souto proposed the idea of synthesis of liposomes; Lucia Ya. Zakharova controlled the physical-chemical part of the work; Oleg G. Sinyashin proposed the idea of the whole work.

**Conflict of interests**

None

**Abbreviations**

AChE, acetylcholinesterase; BBB, blood brain barrier; OP, organophosphorus agent; BChE, butyrylcholinesterase; 2-PAM, Pralidoxime chloride; SNC, central nervous system; DHDAB, dihexadecyldimethylammonium bromide; DHDHAB, dihexadecylmethylhydroxyethylammonium bromide; POX, paraoxon; Ph, L-α-phoshatidylcholine.

**Supplementary material**. Supplementary data are available in the online version, at